\documentclass[12pt,preprint]{aastex}
\shorttitle{Orphaned Protostars}
\shortauthors{Reipurth et al.}

\begin{document}

\title{Orphaned Protostars}

\author{Bo Reipurth\altaffilmark{1}, 
        Seppo Mikkola\altaffilmark{2},
        Michael Connelley\altaffilmark{1},\\
        and
        Mauri Valtonen\altaffilmark{2}}

\vspace{0.5cm}

 \affil{1: Institute for Astronomy, University of Hawaii at Manoa, 
         640 N. Aohoku Place, HI 96720, USA}
  \email{reipurth@ifa.hawaii.edu}
%  \email{msc@ifa.hawaii.edu}

  \affil{2: Tuorla Observatory, 
            Department of Physics and Astronomy,
           University of Turku, V\"ais\"al\"antie
            20, Piikki\"o, Finland}
%\email{Seppo.Mikkola@utu.fi}
%\email{mavalto@utu.fi}

 \newcommand{\simless}{\mathbin{\lower 3pt\hbox
    {$\rlap{\raise 5pt\hbox{$\char'074$}}\mathchar"7218$}}}

\begin{abstract}

  We explore the origin of a population of distant companions
  ($\sim$1000 - 5000~AU) to Class~I protostellar sources recently found
  by Connelley and co-workers, who noted that the
  companion fraction diminished as the sources evolved. Here we
  present N-body simulations of unstable triple systems embedded in
  dense cloud cores. Many companions are ejected into unbound orbits
  and quickly escape, but others are ejected with insufficient
  momentum to climb out of the potential well of the cloud core and
  associated binary. These loosely bound companions reach distances of
  many thousands of AU before falling back and eventually being
  ejected into escapes as the cloud cores gradually disappear. We use
  the term {\em orphans} to denote protostellar objects that are
  dynamically ejected from their placental cloud cores, either
  escaping or for a time being tenuously bound at large separations.
  Half of all triple systems are found to disintegrate during the
  protostellar stage, so if multiple systems are a frequent outcome of
  the collapse of a cloud core, then orphans should be common.  Bound
  orphans are associated with embedded close protostellar binaries,
  but escaping orphans can travel as far as $\sim$0.2~pc during the
  protostellar phase. The steep climb out of a potential well ensures
  that orphans are not kinematically distinct from young stars born
  with a less violent pre-history. The identification of orphans
  outside their heavily extincted cloud cores will allow the detailed
  study of protostars high up on their Hayashi tracks at near-infrared
  and in some cases even at optical wavelengths.

%  After 300,000 yr XXX\% of the triple systems have decayed, and after
%  10~Myr, 90.1\% of the triple systems have decayed.

%  After 10 Myr, the mean velocity of the escapers is 1.13 km/sec, and
%  hence this population does not have a distinct kinematic signature,
%  and does not lead to a diaspora of low-mass ejected objects.

\end{abstract}


\keywords{
binaries: general ---
brown dwarfs ---
stars: formation --- 
stars: low-mass ---
stars: protostars ---
stars: pre-main sequence 
}

\section{INTRODUCTION}

Stars are commonly formed in binary or small multiple systems. This
has been long known from observations (Herbig 1962, Reipurth \&
Zinnecker 1993, K\"ohler \& Leinert 1998), but is also the frequent outcome
of numerical and theoretical work (Goodwin et al. 2007). In fact, a
strong case can be made that the normal outcome of the collapse of a
rotating cloud core is the formation of a binary or multiple system
(Larson 1972, 2002).  It is well known that a system of three bodies
is unstable if they are in a non-hierarchical configuration, and
dynamically always will evolve into either a stable hierarchical
system, or one member will escape and leave behind a bound binary
system (Anosova 1986, Valtonen \& Mikkola 1991). The dynamical and
highly chaotic behavior of multiple systems of young stars has been
extensively explored numerically (Sterzik \& Durisen 1995,1998,
Armitage \& Clarke 1997, Bate et al. 2002, Delgado-Donate et al.
2004). It has been noted that the breakup of a young multiple system
will most often occur during the protostellar stage (Reipurth 2000),
and a consequence of this is that some of the ejected members may not
have gained enough mass to burn hydrogen, thus providing one of the
key pathways for the formation of brown dwarfs (Reipurth \& Clarke
2001, Whitworth et al. 2007).

The most detailed numerical studies so far of a collapsing filamentary
and turbulent cloud leading to a small cluster demonstrate the
enormous dynamical complexity of the many subsystems of stars, with
continuous formation and destruction of multiple stellar systems (Bate
2009).  Although most stars are formed in clusters, in the present
study we focus on the simpler, more tractable problem of the
dynamical behavior of triple systems formed in relative isolation,
without interference from other cluster members, such as found in the
more tranquil environments of stellar associations. We focus in
particular on the ejection of one of the triple members and the
resulting observable consequences.

In recent years, surveys of the binarity of embedded protostars have
begun to appear (Reipurth et al. 2004, Haisch et al. 2004, Duch\^ene
at al. 2004, 2007, Maury et al. 2010).  Connelley et al. (2008a,b)
observed a large number of isolated Class~I sources, and found to
their surprise that these embedded sources have numerous distant
companions. In the projected interval from 963~AU to 4469~AU they found
a clear decrease in the number of companions as function of spectral index
(used as a proxy for age).  In this Letter we
attempt to understand the origin and fate of this population of
loosely bound companions.

%We have determined the fraction of systems in our simulations in which
%a component is observed at radial distances between 1219~AU and
%5657~AU which, after application of a statistical projection factor of
%$\pi$/4, corresponds to 963~AU and 4469~AU in the plane of the sky.

\section{CODE AND CALCULATIONS}

We employ a numerical code for few-body calculations with a
regularization method that provides good accuracy in dealing with the
1/$r^2$ character of the gravitational force as required for a precise
treatment of frequent close triple encounters (Mikkola \& Aarseth
1993). Furthermore, we include the presence of a cloud core
surrounding the young triple system. This gravitational potential has
profound effects on the dynamical behavior of the stars.  We allow the
stars to grow in mass using the prescription of Bondi-Hoyle accretion;
this typically increases the body masses by about 5\%.  Further, we
subtract twice the accreted mass from the core to crudely simulate the
destruction of cloud cores from outflow activity. Finally we let the
remainder of the gas gradually disappear over a period of 300,000 yr,
mimicking the effect of the diffuse interstellar radiation field.

These calculations do not properly represent the dynamical collapse of
the star formation process, during which the bulk of the stellar
masses is rapidly built up in the Class~0 phase when the protostellar
embryos are deeply embedded; this would require a full hydrodynamical
treatment.  Our calculations instead start at the beginning of the
Class~I phase, when the newly formed stars have reached almost their
final masses.  However, we emphasize that the processes discussed here
are also effective during the Class~0 phase, making the disintegration
of newborn multiple systems occur even earlier than we demonstrate
here.  Evans et al. (2009) suggest that, on average, the Class~0 phase
lasts 100,000 yr, and the Class~I phase as long as 440,000 yr.
McClure et al.  (2010) find a shorter embedded lifetime of about
200,000 yr. The figures presented here span an intermediate range of
300,000 yr.

We have performed 12,800 numerical experiments, each spanning 10 Myr.
Three parameters were varied, and 200 simulations run for each
parameter set.  Stellar masses were set to 0.08, 0.2, 0.5, and 1.25
M$_\odot$, representing the most common stellar mass range above the
brown dwarf limit.  Bodies are treated as point masses.  Mean initial
separations were chosen as 50, 100, 200, and 400~AU; for much smaller
separations breakup occurs instantaneously, and for much larger
separations the stars would represent independent star forming events.
The stellar velocities were set so they are approximately virialized,
and the relative initial separations were all non-hierarchical, not
exceeding a ratio of 5:1.  Remnant core masses were set to 0, 1, 3,
and 6 M$_\odot$ with Plummer mass distributions (Plummer 1911), and
radii that remain fixed at 7,500~AU, a size suggested by observations
(Kirk et al.  2006).  The three bodies initially have identical
masses, as the breakup of a system is sped up if one or two of the
bodies are smaller, and we want to take a conservative approach that,
if anything, underestimates the number of escaping bodies.  However,
two of the bodies soon sink towards the center of the core and gain
more mass than the third body (Bonnell et al. 2001), which in most
cases is the body that eventually will escape the system. We do not
consider the angular momentum, if any, of the accreting material onto
the stars, which would have the effect of shrinking the remaining
binaries (Umbreit et al. 2005). The binaries resulting from the
break-up of triple systems would thus be harder than in our
simulations, and consequently we ignore the binaries. The escaping
third bodies, which are the focus of the present study, are, however,
frequently evicted from the interior of the cloud core, and thus spend
much of their time in the tenuous outskirts of the core, less affected
by direct accretion from the core.

\section{RESULTS}

\subsection{Dynamical interactions}

Our 12,800 simulations illustrate the well known chaotic character of
three-body motion. Figure~1 shows four examples of our simulations,
with orbits projected onto the XY-plane. Extinction through the cloud
is indicated by color, black representing an A$_V$ of 50 or higher,
and yellow-white indicating negligible extinction. Local circumstellar
gas is ignored, so the extinction is a lower limit.  However, at least
part of an infalling envelope would be truncated in an ejection event.

Almost immediately, two of the three stars join together in a
(temporary) binary and begin to bounce the third star around. In the
process, it is common that the third star exchanges position with one
of the binary components. Figure~2 contains eight panels showing the
separation of the third body from the binary center of mass as a
function of time, illustrating the rich dynamical behavior of such
systems. Panel~A shows a relatively stable triple system. These are
not common, they are frequently fragile and prone to sudden
disintegration. In contrast, panel~B shows a very common behavior,
where the triple system almost instantaneously breaks up and the third
body escapes. In this case the recoil of the binary is so powerful
that it, too, escapes from the core. However, more frequently the
recoil of the binary is too weak to overcome the gravity of the cloud
core, and the binary oscillates around the center of the core until
the core has dispersed enough to let the binary drift away (Fig.~2{\em
  c}). In yet other cases, both single and binary are immediately
ejected, but with insufficient speed for any of them to escape, so
they rejoin for a period of further close interactions until the
system eventually breaks up (Fig.~2{\em d}). It is common that a
triple system must undergo numerous weaker ejections before an
ejection is finally strong enough to overcome the gradually weakening
potential of the core (Figs.~2{\em e} and {\em f}). Ejections that
almost, but not quite, reach escape speed can lead to giant longlived
excursions spanning many thousands of AU and lasting 100,000~yr or
more before an encounter leads to disintegration (Figs.~2{\em g} and
{\em h}).  Figure~3 shows 100 simulations with color coded ejections
(escapes are red and bound excursions are blue), providing a sense of
the chaotic nature of early triple evolution.

\subsection{Companion fraction as function of parameters}

The companion fraction of a population of young stars is a fairly
easily observed parameter, and we here show that it is influenced by
fundamental properties of the triple systems in the population.

A measurement of a companion fraction must be related to a chosen
range in separation. Figure~4{\em a} shows the companion fraction for four
intervals between 100~AU and 5~pc. Due to the highly dynamic nature
of newborn triple systems, the companion fraction undergoes
significant changes, especially during the protostellar phase. For the
interval 100 to 1,000~AU, almost 100\% of the simulations start with
the most distant component being within this interval. With the rapid decay
of the triple systems, however, the companion fraction in this
interval drops steeply and after 300,000~yr has fallen to about 
10\%. For the interval 1,000 to 10,000~AU, very few simulations start
out with a component in this range, but ejections soon send components
out beyond 1,000~AU. As the escapers move to ever larger distances,
the companion fraction peaks and then diminishes, and is increasingly
accounted for by bound components. For larger separation intervals, a
similar behavior is seen but due to the time required for escapers to
travel to these larger distances, the peak appears at later times and
becomes increasingly broad.

In Figs.~4{\em b}, {\em c}, {\em d} we examine the companion fraction
in the interval 1,000 to 10,000~AU as a function of selected parameters.

In Figure~4{\em b} we see the effect of varying the mass of the cloud
core from 0 to 6~M$_\odot$.  The large number of early ejections
initially boosts the companion fraction, but as the escapers move
away, the companion fraction is increasingly defined by bound systems.
The main effect of the cloud core is to limit the number of ejected
stars that succeed to escape. As a result, the companion fraction
increases with increasing core mass, and for higher core masses almost
stabilizes with time.

The companion fraction turns out, not surprisingly, to be very
sensitive to the mean initial separation of the three bodies
(Fig.~4{\em c}). At 50~AU, the three stars almost instantaneously develop
configurations that decay, leading to a flurry of escapes so the
companion fraction soon drops dramatically. For increasing mean initial
separation, the ejections become weaker leading to fewer
escapes, so the companion fraction increases significantly and only
slowly falls with time as bound systems eventually become dislodged.
After 10 Myr, 10\% of the original 12800 triple systems are still bound,
albeit often tenuously. The fraction of bound systems primarily
depends on the separation: mean initial separations of 50 AU leads to
far fewer bound systems than do 400 AU.

The mass of the stars also affects the companion fraction, as seen in
Fig.~4{\em d}. Perhaps less intuitively, the more massive triple systems
break up more easily, producing an initial peak as escapers pass
through the 1,000 to 10,000~AU interval. For lower masses, the ejected
stars are more readily bound to the core, thus increasing the companion
fraction.

\subsection{Ejection velocity and terminal velocity}

Of the 12,800 simulations performed here, 90.1\% lead to an escape
within the first 10~Myr. The majority of those escapes, however, occur
very early: within the first 200,000~yr already 52\% of the
triple systems have decayed.

The mean ejection velocity of the third body for all ejection events
leading to an escape is 2.8 km/sec, with a tail stretching out to
almost 100~km/sec (Fig.~5).  We sample the velocities approximately
every 1000~yr, and since the peak velocity of the ejection occurs on a
much shorter time-scale, these numbers underestimate the highest
velocities. The peak velocity is attained only very briefly, because
the potential well of the core and binary is steep, and velocities
consequently decline rapidly.

We define the terminal velocity of a population of escapers as the
velocity measured at an age of 10~Myr, independent of the time of
ejection. The mean terminal velocity of all escapers is 1.13 km/sec.
This is well within the turbulent velocity range of molecular clouds,
and escapers are thus kinematically indistinguishable from single
stars born in the same star forming region.

\section{DISCUSSION}

The original motivation for this study was to explore the origin of
the population of distant ($\sim$1000 -- 5000~AU) companions
found around isolated embedded Class~I sources by Connelley et al.
(2008a,b).  Our simulations indeed reveal that the dynamical evolution
of triple systems naturally lead to ejection of components into
distant, longlived excursions, or into escapes. Components that remain
bound are only weakly so and ultimately most will escape, as indeed
observed.  We conclude that the observations are well explained if a
significant fraction of embedded protostars are triple (or
higher-order) systems. We note that these results pertain to isolated
star formation, since stellar interactions in a cluster will break up
wide loosely bound systems (Kroupa 1995, Reipurth et al. 2007, Bate
2009, Parker et al. 2009).

The nature of the escaping components is of considerable interest, and
we here introduce the term {\em orphan} to describe {\em a
  protostellar object which has been dynamically ejected from a
  newborn multiple system}, either into a tenuously bound orbit or
into an escape, thus depriving it from gaining much additional
mass{\footnote{ {\em orphan}: one deprived of some protection or
    advantage [Merriam-Webster]}.  We emphasize that the term orphan
  refers only to {\em protostellar} objects, and thus describes a
  dynamical condition occasionally experienced by a newborn star.  It
  is not easy to say when an orphan no longer is a protostellar
  object, thus ceasing to be an orphan, just as the boundary between
  Class~I and Class~II objects in general is diffuse, but in practical
  terms an object retains its label as an orphan as long as its coeval
  siblings are Class~I objects.  We note a semantic issue here:
  protostars are usually empirically defined by the presence of
  significant circumstellar disks and envelopes. Due to their
  dynamical history, many orphans are likely to have reduced
  circumstellar reservoirs, and thus may sooner lose their
  protostellar characteristics. We use the term protostar here in a
  chronological sense, for objects of an age less than the typical
  duration of the embedded phase (a few hundred thousand years).

  We here summarize the expected properties of orphans:

{\em 1)} Disintegration of triple systems preferentially leads to the
ejection of the lowest-mass member, so orphans are mostly very
low-mass objects, including many brown dwarfs, as discussed by
Reipurth \& Clarke (2001);

{\em 2)} Orphans are often found in distant orbits tenuously bound to
embedded close protostellar binaries, although many are also escaping
shortly after birth;

{\em 3)} Due to their violent dynamical history, orphans are likely to
be surrounded by circumstellar material for a shorter period than if
they had remained in the center of their parental cores (e.g., Clarke
\& Pringle 1993, Bate \& Bonnell 2005), but the youngest will still
display evidence of accretion activity (e.g., emission lines, veiling,
outflows, and/or variability) and all unveiled orphans should display
lithium;

{\em 4)} The amount of circumstellar material that remains bound to an
orphan may vary depending on the specific circumstances of the
ejection process, and this may be a contributing element in explaining
how classical and weak-line T~Tauri stars can have comparable ages but
very different circumstellar environments, as already suggested by
Armitage \& Clarke (1997); 

{\em 5)} Orphans which carry limited circumstellar material, and which
have been ejected out of their nascent cores, may offer a unique
opportunity to study protostellar objects at {\em near-infrared} and
in some cases even at {\em visible} wavelengths;

{\em 6)} Due to the steep climb out of the potential well formed by
the cloud core and the remaining binary, even escaping orphans will
not have unusually high velocities; a consequence of this is that
orphans will not form a diaspora of far-flung stars surrounding a star
forming region (see also Goodwin et al. 2005);

{\em 7)} It is likely that distant third components in hierarchical
triple systems have been ejected in dynamical interactions (Tokovinin
1997, Valtonen 1997, 1998), and those ejected as protostars will have a
pre-history as orphans;

{\em 8)} Orphans will often be found in close association with embedded
Class~0 and Class~I objects, although early escapers may have moved as
much as 0.2~pc (corresponding to $\sim$5~arcmin at the distance of
Taurus) from their site of origin within a few hundred thousand years;

{\em 9)} Orphans are a common result of the breakup of newborn triple
systems; of the 12800 simulations performed here 52\% of the triple
systems disintegrate within 200,000~yr, which is a recent estimate of
the duration of the embedded phase (McClure et al. 2010);

{\em 10)} Most orphans produced in our simulations are single, but it
is not unusual for the remaining binary to temporarily recoil out of
the core center, and in some cases it recoils sufficiently forcefully
to even escape; orphans (including substellar objects) can therefore
also be binaries, and we note that this may be an important source of
brown dwarf binaries;

{\em 11)} Although we expect some orphans to display weaker signatures
of circumstellar material than other protostars, they will -
especially the younger ones - reside high up on their Hayashi tracks,
generally making them more luminous than their more evolved T~Tauri
counterparts (discounting accretion luminosity) that have emerged from their
placental clouds at a more leisurely pace.

{\em 12)} Examples of well-known, probable orphans include T~Tauri
(K\"ohler et al.  2008), HH~111~B (Reipurth et al. 1999), TMR-1C (Riaz
\& Martin 2010), and HBC~515~C/D (Reipurth et al. 2010).

% It is interesting to note the recent discovery of a
% population of young stars in Lupus with curtailed disk reservoirs
% (Comer\'on et al. 2009).  Detection of lithium may be the only
% certain signature of youth in orphan stars.

% * T Tauri systems with IR companions, prototype T Tau itself

% * explains differences of components

% * percent of observed orphans that are escaping

% * no high-velocity orphans (see Goodwin et al. 2005)

% * no strong evidence of youth

% * not a halo of scattered orphans instead well mixed with other YSOs

% * Binaries and the fragmentation scale

\section{CONCLUSIONS}

We have performed N-body calculations of newborn triple systems
embedded in a placental cloud core.  While the members of a triple
system interact in the center of a cloud core they feel only a
fraction of the potential of the core. But as soon as one of the
members is ejected to the outskirts of the core it must climb out of
the full potential well of the core before it can escape. As a result
many fall back, leading to a series of ejection events, and thus a
yo-yo-like motion of the star takes place. To escape, the star must
await either a particularly forceful close triple encounter or wait
until the core has lost enough mass to lower its potential barrier.
We conclude that the loosely bound distant companions to embedded
protostars found by Connelley et al. (2008a,b) are well explained as
orphans.  The identification and observational study of orphans may
offer important insights into the very earliest stellar evolutionary
stages.

\acknowledgments

We thank Hans Zinnecker for valuable discussions.
This work was supported by the NASA Astrobiology Institute under
Cooperative Agreement No. NNA04CC08A.

% REFERENCES

\clearpage 

% NOTE THAT USING COLOR FIGURES COSTS $350 PER ARTICLE, NOT PER PAGE!

\clearpage

%USE {figure*} EVERYWHERE BELOW WHEN USING THE TWO-COLUMN FORMAT

\begin{figure}
\epsscale{1.0}
\plotone{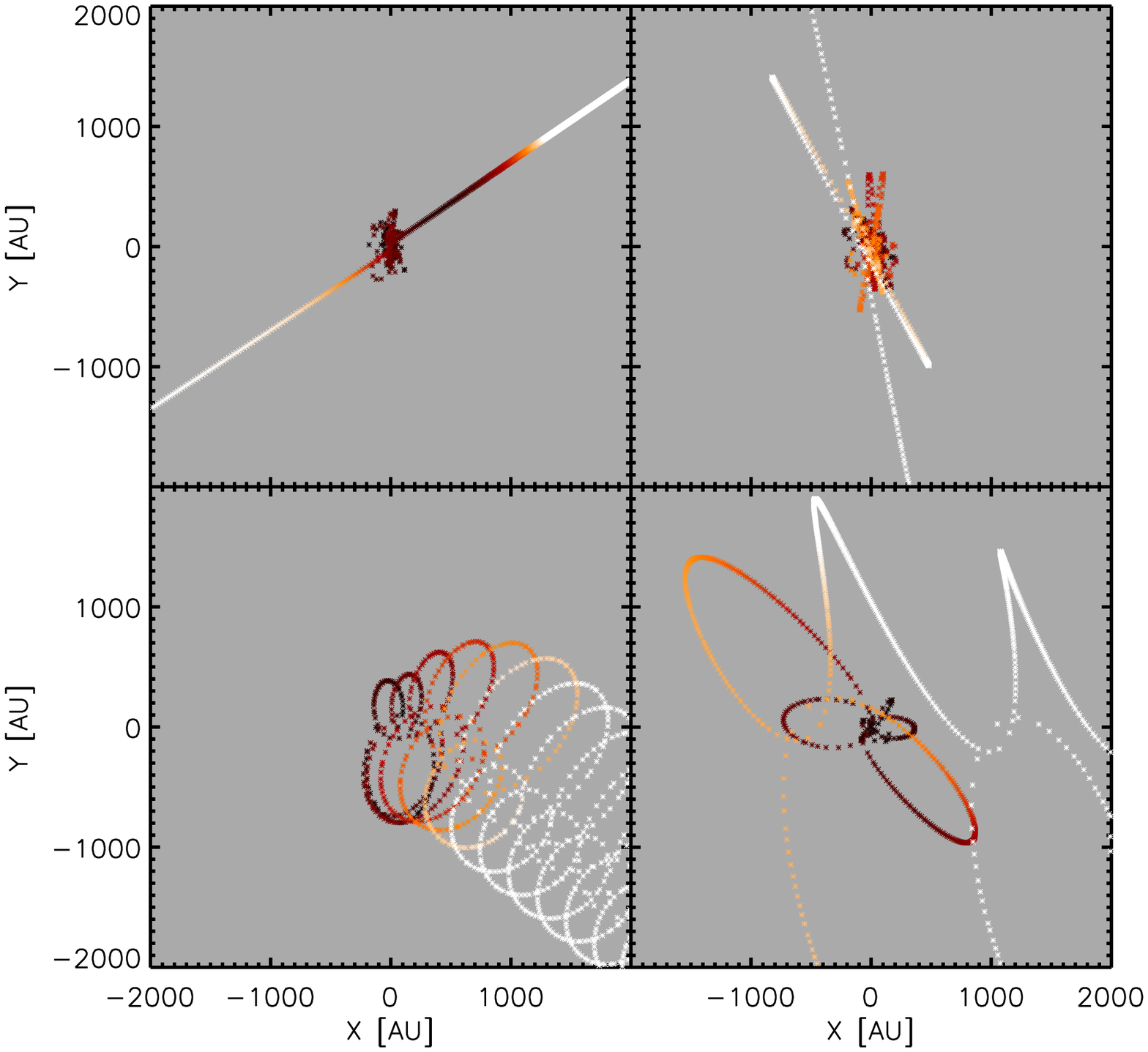}
\caption{ Four examples of physical orbits projected on the XY-plane.
  Color indicates extinction from the cloud core, from black
  representing A$_V$=50 or higher down to light yellow/white
  representing little or no extinction. The core dissipates after
  300,000~yr, so the stars become visible with time or as they are
  ejected or drift out of the core.  \label{fig1}}
\end{figure}

\begin{figure}
\epsscale{1.0}
\plotone{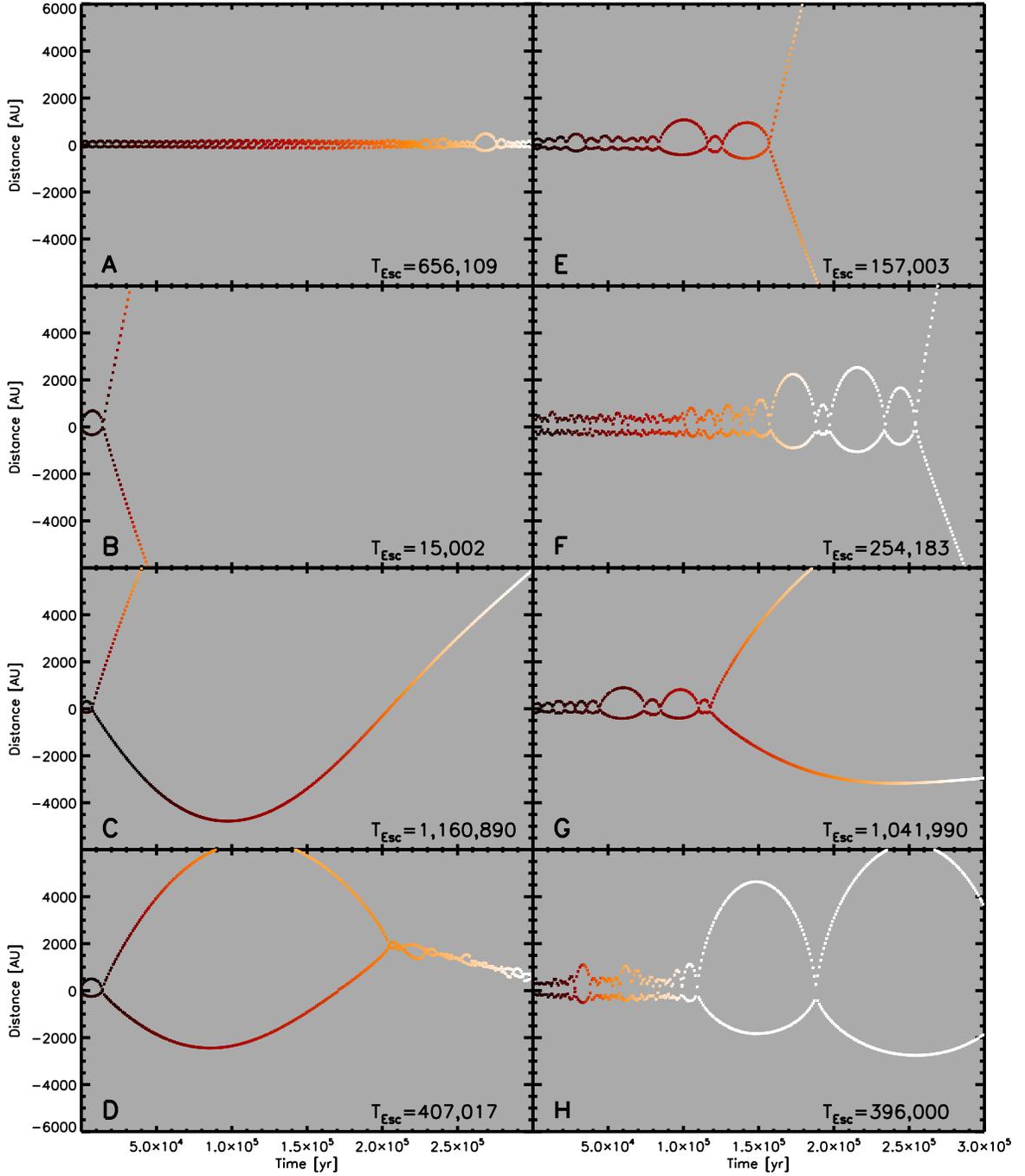}
%\plotone{fig2mosaic4_old.eps}
%\plotone{fig2b.ps}
%\plotone{fig2c.ps}
%\plotone{fig2d.ps}
\vspace{-2.0cm}
\caption{Samples of characteristic dynamical behavior of single and
  binary components of an unstable triple system. The single is always
  towards the top, and the binary towards the bottom.  Color indicates
  the extinction, from black representing A$_V$=50 or higher down to
  light yellow/white representing little or no extinction. The
  line-of-sight is along the Z-axis of the simulations.  Numerous
  ejections are occurring, but only three cases (B, E and F) lead to
  an escape during the time interval shown. Eventually all the cases
  shown lead to an escape (that is, single and binary meet again, even
  in case C) and the time of escape is listed for each.  \label{fig2}}
\end{figure}

\clearpage

\begin{figure}
\epsscale{1.0}
\plotone{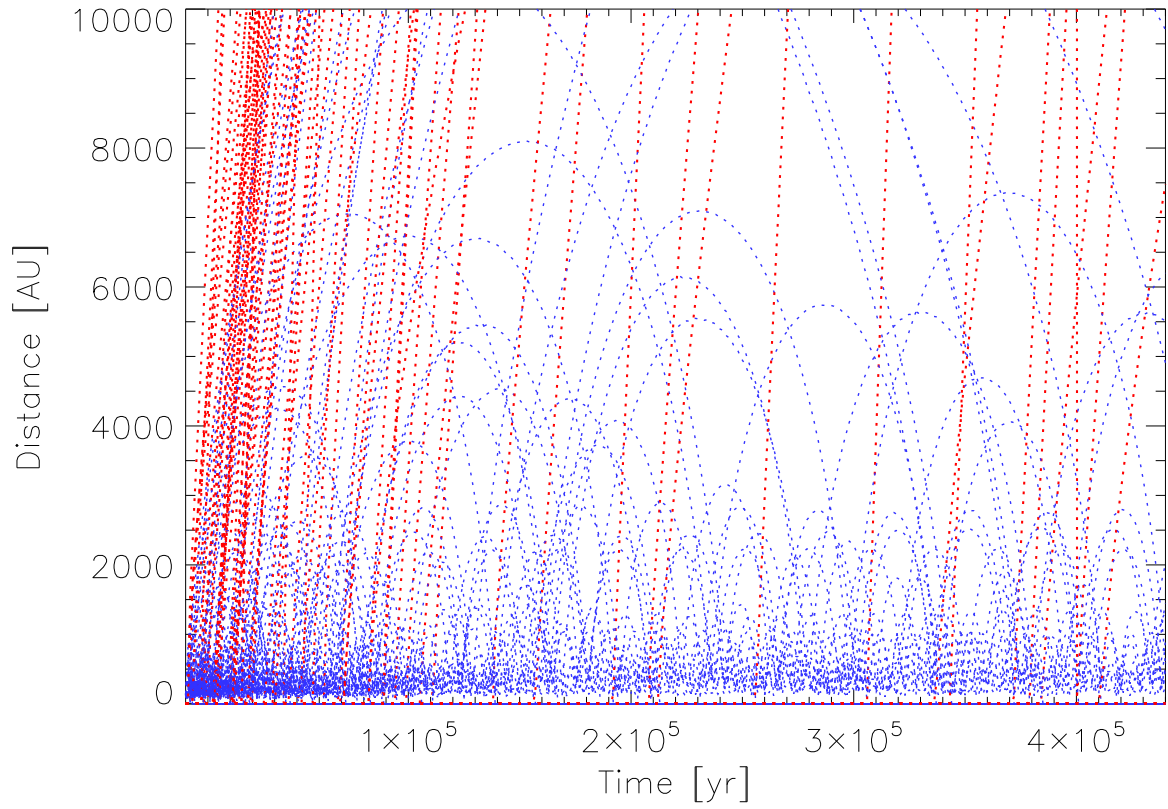}
\caption{100 simulations showing the dynamical evolution of a triple
  system of three 0.5~M$_\odot$ stars with initial mean separations of
  100~AU embedded in a 3~M$_\odot$ cloud core. Among the numerous
  ejections seen in the plot, those leading to an escape are plotted
  in red, while bound systems are blue. \label{fig3}}
\end{figure}

\clearpage
    
\begin{figure}
\epsscale{0.45}
\plotone{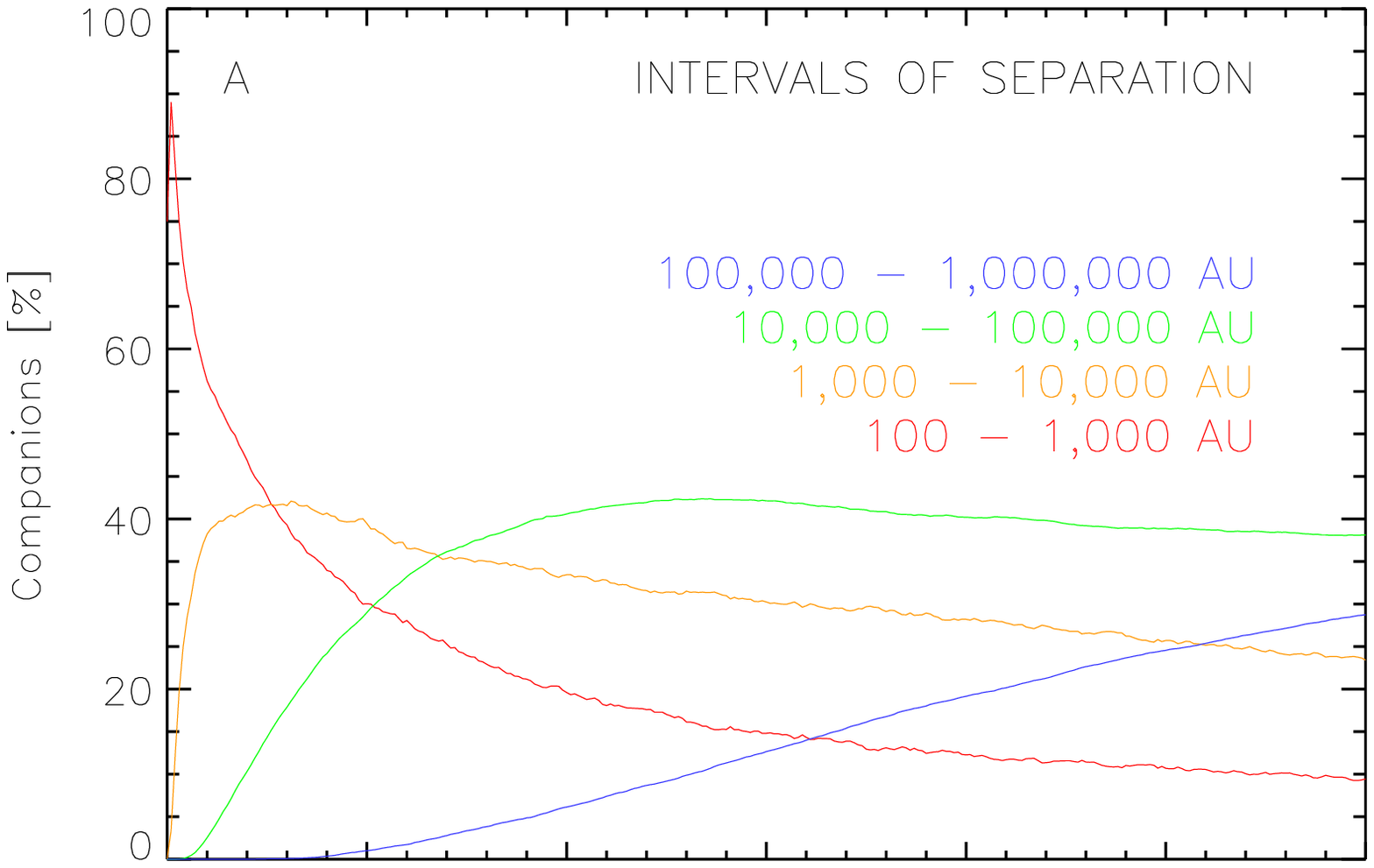}
\plotone{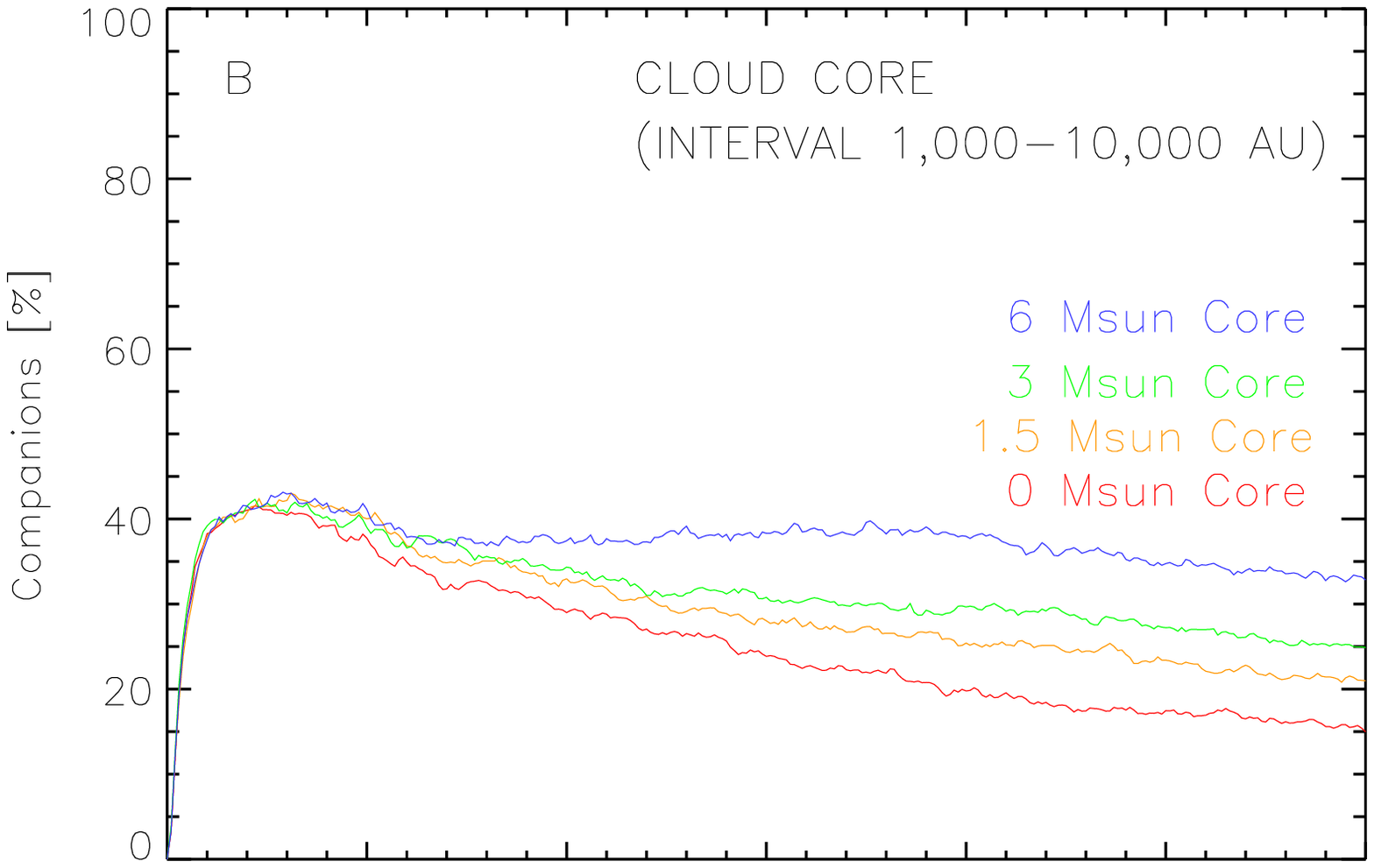}
\plotone{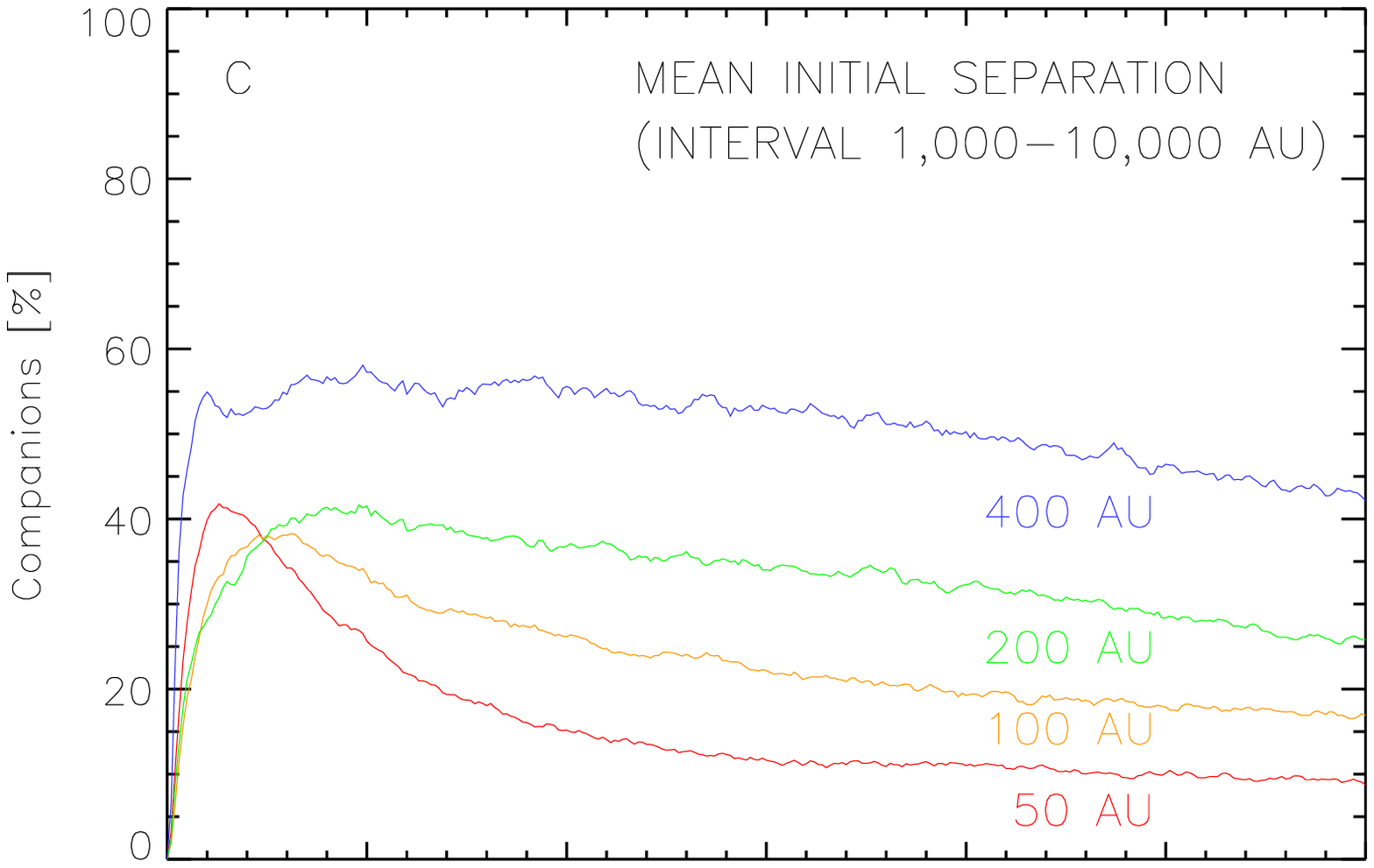}
\plotone{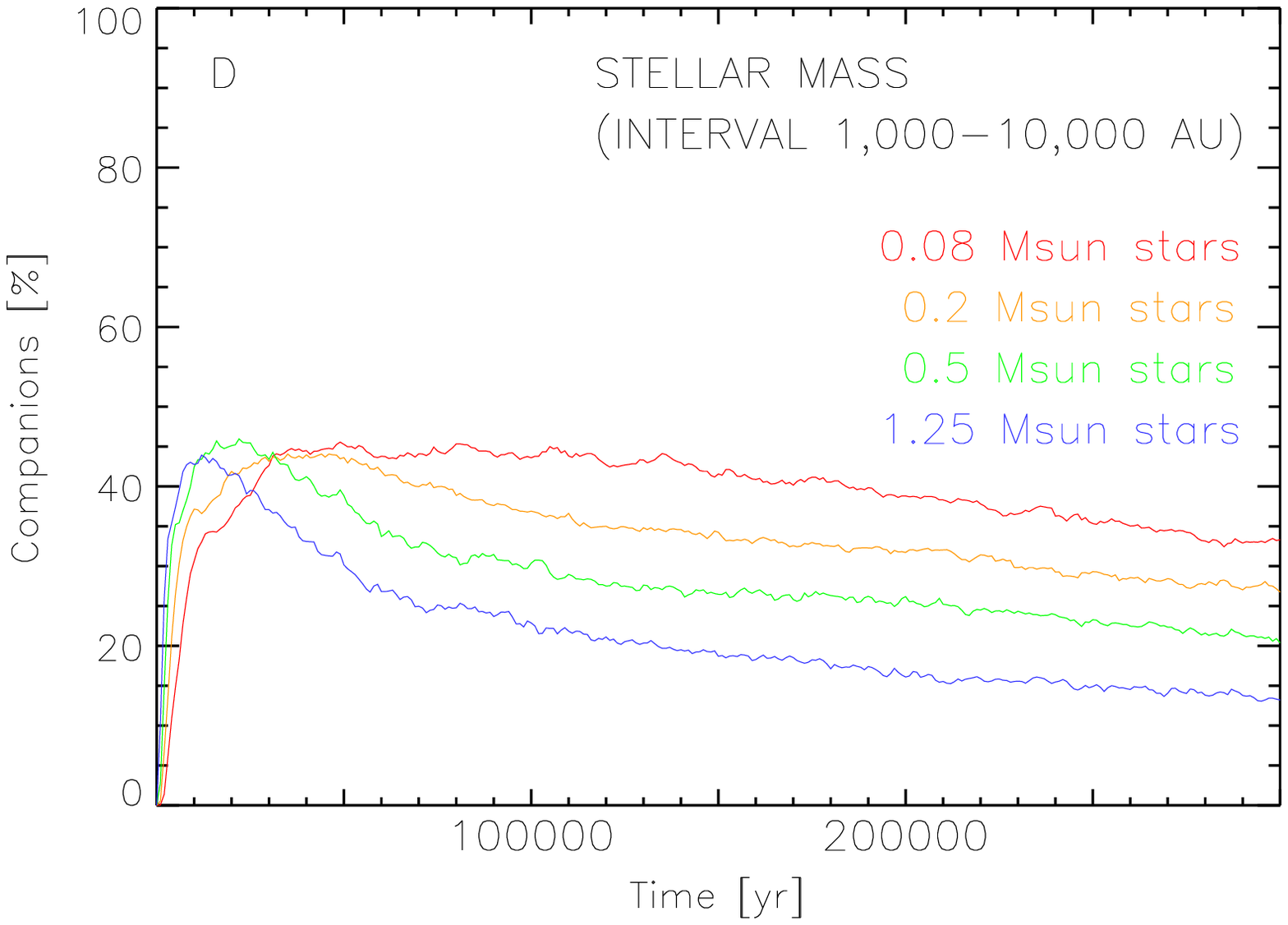}
\caption{The companion fraction as function of various parameters. 
In panel~A the companion fraction is plotted as function of time in 
four different intervals of separation. In panels~B,~C,~D the 
effect of varying core mass, mean initial separation, and stellar mass
is shown for the interval 1,000 to 10,000~AU. Mean of all 12800 simulations.
\label{fig4}}
\end{figure}

\clearpage

\begin{figure}
\epsscale{1.0}
\plotone{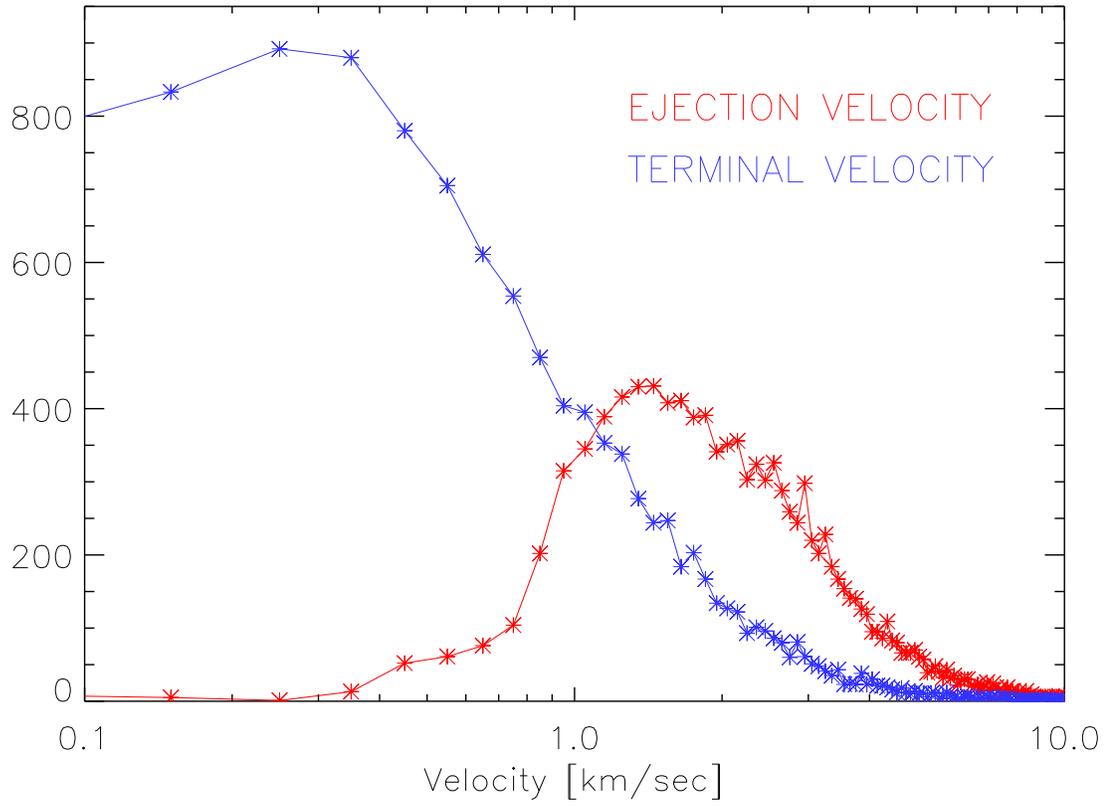}
%\plotone{ev_ALL80_Part4.ps}
%\plotone{tv_ALL80_Part4.ps}
\caption{The distribution of ejection and terminal velocities 
plotted in 0.1~km/sec intervals. For the terminal velocity the large majority of cases are found below 2~km/sec.  
\label{fig5}}
\end{figure}

\end{document}